\documentclass[12pt]{iopart}

\usepackage{graphicx} 
\usepackage{dcolumn}  
\usepackage{bm}       
\usepackage{amssymb}
\newcommand{\ba}{\begin{eqnarray}}
\newcommand{\ea}{\end{eqnarray}}
\newcommand{\kB}{k_{_B}}

\newcommand{\rhohac}{\rho_c^{\hbox{\tiny{(h)}}}}
\newcommand{\nha}{n^{\hbox{\tiny{(h)}}}}

\newcommand{\sha}{s^{\hbox{\tiny{(h)}}}}
\newcommand{\shac}{s_c^{\hbox{\tiny{(h)}}}}
\newcommand{\shacth}{s_c^{\hbox{\tiny{(h)}}}|_{_{\hbox{\tiny{th}}}}}

\newcommand{\shacem}{s_c^{\hbox{\tiny{(h)}}}|_{_{\hbox{\tiny{em}}}}}
\newcommand{\nhac}{n_c^{\hbox{\tiny{(h)}}}}
\newcommand{\xhac}{x_c^{\hbox{\tiny{(h)}}}}
\newcommand{\Tha}{T^{\hbox{\tiny{(h)}}}}
\newcommand{\xha}{x^{\hbox{\tiny{(h)}}}}
\newcommand{\sigha}{\sigma_{\hbox{\tiny{(h)}}}}
\newcommand{\xf}{x_{\hbox{f}}}
\newcommand{\nf}{n_{\hbox{f}}}
\newcommand{\sff}{s_{\hbox{f}}}
\newcommand{\tb}{\tan \beta}
\begin{document}

\title[Constraining the mSUGRA parameter space using the entropy of dark matter halos]{Constraining the mSUGRA parameter space using the entropy of dark matter halos}

\author{Dar\'\i o N\'u\~nez$^1$, Jes\'us Zavala$^1$\footnote{Present address: Shanghai Astronomical Observatory, Nandan Road 80, Shanghai 200030, China; jzavala@shao.ac.cn}, Lukas Nellen$^1$, Roberto A. Sussman$^1$\footnote{Present address: Instituto de Fisica, Universidad de Guanajuato, Loma del Bosque 103, Leon, Guanajuato, 37150, Mexico (on sabbatic leave from ICN-UNAM)}, Luis G. Cabral-Rosetti$^2$, and Myriam Mondrag\'on$^3$ (For the Instituto Avanzado de Cosmolog\'{\i}a, IAC)}

\address{$^1$ Instituto de Ciencias Nucleares, Universidad Nacional Aut\'onoma de M\'exico (ICN-UNAM) A. P. 70-543,  M\'exico 04510 D.F., M\'exico}

\address{$^2$ Departamento de Posgrado, Centro Interdisciplinario de Investigaci\'on y Docencia en Educaci\'on T\'ecnica (CIIDET), Av. Universidad 282 Pte., Col. Centro, A. Postal 752, C. P. 76000, Santiago de Quer\'etaro, Qro., M\'exico.}

\address{$^3$ Instituto de F\'{\i}sica, Universidad Nacional Aut\'onoma de M\'exico (IF-UNAM), A. Postal 20-364, 01000 M\'exico D.F., M\'exico.}

\eads{\mailto{nunez@nucleares.unam.mx}, \mailto{jzavala@nucleares.unam.mx}, \mailto{lukas@nucleares.unam.mx}, \mailto{sussman@nucleares.unam.mx}, \mailto{lgcabral@ciidet.edu.mx} and \mailto{myriam@fisica.unam.mx}}

\begin{abstract}

  We derive an expression for the entropy of a dark matter halo
  described by a Navarro-Frenk-White model with a core. The comparison
  of this entropy with the one of dark matter at the freeze-out
  era allows us to constraint the parameter space in mSUGRA models.
  Moreover, joining these constraints with the ones obtained from the
  usual abundance criteria and demanding both criteria to be
  consistent with the 2$\sigma$ bounds for the abundance of dark
  matter: $0.112\leq\Omega_{DM}h^2\leq0.122$, we are able to clearly
  discriminate validity regions among the values of $\tan \beta$, one
  of the parameters of the mSUGRA model. We found that for the
    explored regions of the parameter space, small values of
  tan$\beta$ are not favored; only for tan$\beta\simeq50$ are both
  criteria significantly consistent. In the region where both criteria
  are consistent we also found a lower bound for
  the neutralino mass, $m_{\chi}\geq 141$ GeV. 
\end{abstract}

\noindent{\it keywords\/}: dark matter, cosmology of theories beyond the SM, structure of galaxies


\section{Introduction}

The existence of components of the total energy-density of the Universe
whose nature is still not understood, constitutes one of the biggest unsolved
questions in physics today. It is a great challenge 
to have a clear picture of the nature of dark matter (DM), and dark energy, whose
existence is more and more undisputed with the observational evidence accumulated
over the last decades.

The observational constraints on the present density of DM,
$\Omega_{DM}$, come from several outstanding observations such as the
Cosmic Microwave Background radiation (CMBR) \cite{WMAP}, galaxy
clustering, supernovae and Lyman $\alpha$ forest. One of the most
recent works which combines all these data leads to:
$0.112\leq\Omega_{DM}h^2\leq0.122$ \cite{seljak}, with $h$ the
dimensionless value of the present day Hubble constant ($h \sim 0.7$).
These bounds imply that dark matter constitutes $22.8 - 24.8\,\%$ of
the total density of the Universe.

One of the most accepted candidates to be the major component of dark matter 
is the neutralino as the Lightest Supersymmetric Particle (LSP).
Supersymmetric theories with R-parity conservations predict this 
particle (for an excellent introduction to Supersymmetry see \cite{martin}).
This type of models have several parameters, some of which can be constrained by combining different
methods with observational data of the actual density of DM. 
In particular for mSUGRA models, this has been done using a standard approach 
\cite{belanger,constraint}, based on solving the Boltzmann
equation by considering that after the ``freeze-out'' era, neutralinos cease
to annihilate keeping its number constant. In such approach, the relic
density of neutralinos is approximately given by (this is strictly true only when $\langle\sigma v\rangle$
is independent of the energy): 
\begin{equation}
\Omega_{\chi}\propto 1/\langle\sigma v\rangle, \label{OmegaAC}
\end{equation}
where $\langle\sigma v\rangle$ is the thermally
averaged cross section times the Moller velocity of the annihilating
pair.

Within the mSUGRA model, five parameters are needed to specify the
supersymmetric spectrum of particles and the final relic density of
the LSP. These parameters are: $m_0$, the unified mass for scalars,
$m_{1/2}$, the unified mass for gauginos, $A_0$, the unified trilinear
coupling, tan$\beta$, the ratio of the vacuum expectation values of
the neutral components of the supersymmetric Higgs bosons and the sign
of $\mu$, where $\mu$ is the Higgsino mass parameter.  In this work we
use the numerical code micrOMEGAs \cite{micro} to compute the LSP's
relic density for different values of these parameters. This scheme
will be called hereafter the ``abundance criterion'' (AC), which
combined with the hypothesis $\Omega_{DM}=\Omega_{\chi}$, gives an
effective way to constraint the values of those five parameters.

Another estimate for $\Omega_{\chi}$ can be obtained using a different
approach.  Just before the ``freeze-out'' epoch, we can consider
neutralinos as forming a Maxwell-Boltzmann (MB) gas in thermal
equilibrium with the rest of the components of the Universe. In the
present time, such a gas is almost collisionless and either
constitutes galactic halos and larger structures or it is in the
process of their formation. In this context, we can conceive two
equilibrium states for the neutralino gas, the decoupling (or
``freeze-out'') epoch and its present state as a virialized system.
Computing the entropy per particle for each one of these states, we
can use an ``entropy consistency'' criterion (EC), based on the
comparison of theoretical and empirical estimates for this entropy, to
obtain an alternative expression for the relic density of neutralinos
$\Omega_{\chi}$, functionally dependent of the mSUGRA parameters. This
idea was originally introduced by some of us in \cite{lyr}.

Our objective is then to develop the recently introduced EC, as a complementary method to the well known AC,
in order to obtain further constraints on the parameters of the mSUGRA model by demanding 
consistency of both criteria with each other and with the observational 
constraints on $\Omega_{DM}$ (preliminary results of this analysis were already presented in \cite{proceeding}). 

The paper is organized as follows, in section II we present a brief
description of the method to compute $\Omega_{\chi}$ using the AC. In section
III a derivation of the EC is explained. In section IV we present the halo model together
with a detailed description of the method we followed to obtain one of the key empirical parameters
in the EC. In section V we
present the results obtained by comparing both criteria with the
observational constraints on $\Omega_{DM}$ in the context of the mSUGRA model. In the last section a summary and
the conclusions of our work are presented. 

\section{Abundance criterion}

The relic abundance of neutralinos is defined as 
$\Omega_{\chi}=\rho_{\chi}/\rho_{crit}$, where $\rho_{\chi}=m_{\chi}n_{\chi}$
is the relic's mass density ($n_{\chi}$ is the number density, $m_\chi$ the neutralino mass) and $\rho_{crit}$
is the critical density of the Universe (see \cite{kamion} for a review
on the following method to compute the relic density). The time evolution of
$n_{\chi}$ is given by the Boltzmann equation:
\begin{equation}\label{bolt}
\frac{dn_{\chi}}{dt}=-3Hn_{\chi}-\langle\sigma v\rangle(n_{\chi}^2
-(n_{\chi}^{eq})^2), \label{eq:Bol}
\end{equation}
where $H$ is the Hubble expansion rate and $n_{\chi}^{eq}$ is the number density of neutralinos
in thermal equilibrium. In the early Universe, neutralinos 
and the rest of species were in thermal equilibrium, that is $n_{\chi}=n_{\chi}^{eq}$. As the 
Universe expanded, their typical interaction rate started to diminish and the
process of annihilation froze out. Since then, the global comoving number density of neutralinos has
remained nearly constant.

There are several ways to solve Eq.~(\ref{bolt}). The most common approach
is based on the ``freeze-out'' approximation (see for example \cite{gondolo}). 
If we consider that neutralinos are the major component 
of the dark matter density, $\Omega_\chi\approx\Omega_{DM}$, then the freeze-out approximation gives a
functional dependence between $\Omega_{DM}$ and the mSUGRA parameters. In other words, it gives a
constraint equation for these parameters.

In practice, however, we obtain a numerical solution to the
Boltzmann equation, Eq.~(\ref{eq:Bol}), using the public code micrOMEGAs 1.3.6  
\cite{micro}, this gives us better precision over the usual freeze-out approximation. 
The code computes the relic density of the LSP in the Minimal 
Supersymmetric Standard Model (MSSM), taking into account all annihilation and 
coannihilation processes as well as loop-corrected masses and mixings to calculate
$\langle\sigma v\rangle$ exactly. Further on, we take the mSUGRA model and its 
five parameters ($m_0$, $m_{1/2}$, $A_0$, tan$\beta$ and the sign of $\mu$) 
as input for micrOMEGAs and use {\it Suspect} in its version 2.34 \cite{suspect},
which comes as an interface to micrOMEGAs, to calculate the supersymmetric 
mass spectrum of particles. 

Using this criterion, we can obtain the allowed regions in the parameter space of the
mSUGRA model which are consistent with observational constraints on the actual dark matter density.

\section{Entropy consistency criterion}

This criterion was originally introduced in \cite{lyr}. We briefly describe it in the following.

The main idea of the method is to compare the neutralino gas in two
stages of its evolution as the Universe evolved, namely, the freeze-out era and the present epoch.
These initial and final states are taken as equilibrium states. In this 
approach we disregard the complex phenomena that took place during the formation
and evolution of structure in the Universe, and the analysis is focused only on those
two stages. In the freeze-out era, the assumption is to treat the neutralino gas
as a non-relativistic ideal gas of WIMPs (weakly interacting massive particles) 
described by Maxwell-Boltzmann statistics. This assumption is valid mainly because in this era, the dominant 
interactions in the gas are still
short-ranged and neutralinos are already non-relativistic by then ($m_{\chi}\gtrsim 100$ GeV).

We would like to apply the MB description for present day structures as well, in particular
to galactic halos, which we consider as final states of the primordial neutralino gas.
However, the valid assumptions made for the
freeze-out era are no longer valid for such final states, because in galactic halos,
neutralinos are mainly subject to long-range interactions, related to non-extensive forms
of energy and entropy. Following the guidelines given in \cite{lyr}, a more general approach
can be applied instead, it uses the microcanonical ensemble in the ``mean field'' approximation.
This approach is actually valid at both, the initial and final stages. The state of the system in
each state is defined by its entropy per particle s, its density n, and its temperature T; the
latter is actually given by the auxiliary variable $x=\frac{m_\chi}{T}$. This set of variables
is then ($\sff,\,\xf,\,\nf$) for the neutralino gas at the freeze-out era and ($\sha,\,\xha,\,\nha$)
for the dark matter halo today. Therefore, the change in entropy per 
particle between these two states is given by:
\ba \sha \ - \ \sff \ = \ 
\ln\,\left[\frac{\nf}{\nha}\left(\frac{\xf}{\xha}\right)^{3/2}
\right].\label{Delta_s}\ea
The region today in which all these considerations apply is the 
center of galactic halos. Thus, in what follows, we consider 
current halo macroscopic variables as evaluated in this region:
$\shac,\,\xhac,\,\nhac$.

We can then write $\shac$ using Eq.~(\ref{Delta_s}) as an equation
depending only on present day values of certain quantities, such as the ratio of the total
density of the Universe to $\rho_{crit}$, $\Omega_0$, and the value 
of the freeze-out temperature of the neutralino gas. This was done in detail
in \cite{lyr} and the result can be written as: 
\begin{equation}\label{shalo}
\shacth = \frac{5}{2}+\xf+\ln\left[\frac{g_{*\hbox{f}}\left(\xf\right)\left(x_0^{\hbox{\tiny{CMB}}}\right)^3}
{g_{*0}\left(x_0^{\hbox{\tiny{CMB}}}\right)}\frac{h^2\,\Omega_0}
{(\xf\,\xhac)^{3/2}}\,\frac{\rho_{crit}}{\rhohac}\right].
\end{equation}
where $g_*$ is related to the degrees of freedom of the system, 
$x_0^{\hbox{\tiny{CMB}}}\equiv m_\chi/T_0^{\hbox{\tiny{CMB}}}=4.29\times10^{12}\,m_\chi/\hbox{GeV}$, with 
$T_0^{\hbox{\tiny{CMB}}}=2.7\,\hbox{K}$ (see \cite{gondolo} for details).

It is important to mention that there is an alternative formalism to study
autogravitating systems such as dark matter halos. It is based on a non-extensive description
of energy and entropy developed by Tsallis \cite{Tsallis}. Such formalism was applied
to dark matter halos and compared with the predominant $\Lambda$CDM paradigm
obtaining suggestive results \cite{Tsallis2p, Tsallis2}. However, for the purposes of this work 
we will continue with the traditional formalism.

If the neutralino gas in present halo structures would strictly satisfy MB 
statistics, the entropy per particle would follow from  the well known 
Sackur--Tetrode entropy formula \cite{Pathria}. Such a MB gas in equilibrium 
would be equivalent to an isothermal halo if we make the connection \cite{BT}: 
$\sigha^2=\kB\Tha/m$,
where $\sigha^2$ is the velocity dispersion (a constant for isothermal halos). 
However, as mentioned before, the assumption of MB statistics does not apply for 
self--gravitating collisionless systems. Hence, an exactly isothermal halo is 
not a realistic model, not only because of these theoretical arguments, but 
also because its total mass diverges and its distribution function allows 
for infinite particle velocities (theoretically accessible in the velocity 
range of the MB distribution). 

More realistic halo models follow from ``energy truncated'' distribution 
functions \cite{BT}-\cite{MPV}\nocite{Padma3,Katz1,Katz2} that assume a maximal ``cut off'' 
velocity (an escape velocity). Therefore, we can provide a convenient 
empirical estimate of the halo entropy, $\shac$, following the microcanonical 
definition of entropy in terms of the allowed phase space volume, restricted 
to the range of velocities accessible to the central 
particles, that is up to a maximal escape velocity $v_e(0)$. If we reasonably assume  
that:
\ba 
v_e^2(0) \ = \ 2\,|\Phi(0)| \ \simeq \ \alpha \, \sigha^2(0),
\label{alphas}
\ea
where $\Phi(r)$ is the Newtonian gravitational potential, and $\alpha$ is a
proportionality constant, then we can write:
\begin{equation} 
\shacem \ \simeq\ln\left[\frac{m^4\,v_{e}^3}{(2\pi\hbar)^3\,\rhohac}\right]\nonumber=89.17
+\ln\left[\left(\frac{m}{\hbox{GeV}}\right)^4\,
\left(\frac{\alpha}{\xha_c}\right)^{3/2}
\,\frac{\hbox{GeV/cm}^3} {\rhohac}\right]
\label{SHALO}
\end{equation}
where we used $\xhac=c^2/\sigha^2(0)$ with $c$ the speed of light.

Equating the theoretical and empirical estimates for the entropy per particle (Eqs. (\ref{shalo}) 
and (\ref{SHALO})) we finally obtain 
a relation for the relic abundance of neutralinos (this formula is a small modification 
of the one presented in \cite{lyr}):
\begin{equation}\label{OmegaEC}
\textrm{ln}(\Omega_{\chi}h^2)=10.853-\xf+\textrm{ln}\left[\frac{(\xf\alpha)^{3/2}m_{\chi}g_{*0}\left(x_0^{\hbox{\tiny{CMB}}}
\right)}{g_{*\hbox{f}}\left(\xf\right)}\right] 
\end{equation}
In this way we have obtained another constraint equation for the parameters of the mSUGRA model (recall that 
$\Omega_{DM}=\Omega_\chi$).

In order to perform an analysis of the mSUGRA parameters using this new entropy criterion, we
modified micrOMEGAs to compute the value of $\xf$ for any
region of  the parameter space, and then compute $\Omega_{\chi}$ using Eq.~(\ref{OmegaEC}).
We can then find the subspace of values of the mSUGRA parameters consistent with the observed
constraints on the dark matter density.

The extra parameter $\alpha$ in Eq.~(\ref{OmegaEC}), which is the proportionality factor 
between the escape and dispersion velocities at the halo center has a major role in our work. 
It is very important for the EC that we are able to give values to $\alpha$ in an independent way. 
In the following section, we
show in detail how this parameter can be evaluated for a particular dark matter halo model. 

The scheme we have developed so far is only for dark matter particles, namely neutralinos. We have not
included at any moment the baryons which constitute the galaxies that we observe today. As we discuss 
further on, this condition makes our method strictly valid only for pure dark matter structures and gives
a reasonable approximation when the dynamical effects of baryons can be neglected. The main effects coming
from the addition of baryons that should be included in our model are: i) neutralino-nucleon 
elastic scattering interactions and ii) gravitational effects from the baryonic component onto the dark 
matter halo. Although both of these effects are more significant in the central region of the halos, precisely where our
method should be applied, we believe that the latter can be avoided if we focus our analysis on systems where 
the baryon-to-dark mass content is significantly low, and the former can be safely neglected in all cases 
since the interaction rate between neutralinos and baryons is highly suppressed by the low neutralino-proton 
scattering cross section predicted by theoretical models $~10^{-44}-10^{-45}$ cm$^2$ (see for example
\cite{Feng}, see also \cite{CDMS} for recent experimental upper limits). The formation of 
a baryonic component in the galactic (halo+baryons) system, a disc for example, modifies the original mass distribution of 
dark matter, the effect is a net contraction of the halo towards the center, see for example \cite{Mo} for
a classical description on how to include this effect. In what follows we will not consider it though, since
we will focus on dark matter dominated systems where the effect is less significant.

\section{About the parameter $\alpha$}

The formula for the entropy per particle in the center of halos, 
Eq.~(\ref{SHALO}), is far from being the final description
from a statistical-mechanics point of view. The question about the actual description 
governing ``dark matter fluids" remains open. Nevertheless, it is the assumption we 
have made about the neutralino gas, taking it
nearly as an ideal gas, that make Eq.~(\ref{SHALO}) consistent. In fact,
other works in the past have considered the approximation of dark matter as an ideal gas as the 
approach to follow when it comes to analyzing the entropy (see for example \cite{eke_ent}, 
\cite{entropia_halos}). Our ignorance on the 
correct statistical-mechanics treatment for systems formed by dark matter is
reflected in the appearance of the parameter $\alpha$ in Eq.~(\ref{SHALO}). It is of key importance
to find appropriate bounds for its  value.
In the following, we present an specific model used to obtain estimates for such bounds. 
We have chosen this model trying to balance simplicity with an 
approximate general description of dark matter halos.

The model consists of an spherical dark matter halo with a constant density core in the center,  
beyond which the dark matter density profile follows the well known Navarro-Frenk-White (NFW) 
profile \cite{NFW}, we set a cut-off radius to define the halo boundary:
\begin{displaymath}
\rho(y)=\left\{\begin{array}{ll}
\rho_c & \textrm{if $y < y_c$}\\
\frac{\delta_0\rho_0}{y(1+y)^2} & \textrm{if $y_c \leq y \leq y_v$}\\
0 & \textrm{if $y>y_v$}
\end{array} \right.
\end{displaymath}
\begin{equation}
\,\ \label{rho}
\end{equation}
where $\rho_c$ is the constant central density of the core, $y=r/r_s$, $y_c=r_c/r_s$, $y_v=r_v/r_s$, $r_s$ is a scale 
radius, $r_c$ is the core radius and $r_v$ is the virial radius, which defines the limit for
the virialized structure; $\rho_0$, $\delta_0$ and $r_s$ 
are the usual parameters of the the NFW profile. The choice of an NFW profile for the
external part is clearly motivated by the success of the profile as a universal fit to 
dark matter halos in numerical simulations. The central core is motivated by the apparent
need of it in dark matter dominated systems, see further below, and also by the phenomenon of 
dark matter annihilation which could produce it naturally (see for example \cite{Nata07}).

For a model without core, all these parameters can be given by a series of
well-established formulas (see for example \cite{Tsallis2,NFW, Mo, Lokas}):
\begin{eqnarray} 
\delta_0=\frac{\Delta\,c^3}{3\left[\ln\,(1+c)-c/
(1+c)\right]},\label{delta0}\\\nonumber\\
\rho_0=\rho_{\mathrm{crit}}\,\Omega_0\,h^2 = 277.8 \, h^2\,
\frac{M_\odot}{\rm{kpc}^3},\label{rho0}
\end{eqnarray}
where $c=r_s/r_v$ is the concentration parameter. We choose a $\Lambda$CDM model
with $\Omega_0=1, \Omega_{\Lambda}=0.7$, which seems to be in accordance with
different sources of observations \cite{WMAP}; and 
$\Delta\sim 100$ \cite{Lokas2,Lokas3}.

With Eqs.~(\ref{delta0}) and (\ref{rho0}) and the definition of the
parameter $c$, the NFW density profile is defined by two parameters only,
a ``size'' parameter ($r_s$ or $r_v$) and the concentration parameter $c$. In fact, both
can be written in terms of the total mass of the halo, $M_{v}$:
\begin{eqnarray} 
r_v=\left(\frac{3M_v}{4\pi\Delta\rho_0}\right)^{1/3},\label{r_v}\\\nonumber\\
c_0\approx160\left(\frac{M_vh}{M_{\odot}}\right)^{-0.096}.\label{c0}
\end{eqnarray}
Eq.~(\ref{c0}) is a fit extracted from the numerical 
study \cite{Neto}, which is based on the Millennium simulation data \cite{MS}, the sample of relaxed
halos presented in that work was chosen as more appropriate for our purposes (the halos in this sample are
closer to virial equilibrium). The
results were transformed from the original value $\Delta=200$ to $\Delta=100$,
to do so, we used the code provided by the authors.
Other studies in the past have found different fitting formulas for the mass-concentration relation, we believe 
our results are
not highly sensitive to a specific formula (we actually tested this with the formula used in\cite{Zavala}). 
Eq.~(\ref{c0}) however, just gives a central value for the 
concentration (this is why we use the term $c_0$ to represent it), actually, the
relation between $c$ and $M_v$ in numerical studies has a scatter, 
$\sigma_{log c}$, which is slightly dependent on $M_v$ itself. For the remainder of this work, we take however the same 
  value for all halos, given by the mean of $\sigma_{log c}$ for the sample of relaxed halos presented in \cite{Neto} 
  (see table 1 of their paper): $\sigma_{log c}\approx 0.095$.

Using Eqs.~(\ref{delta0}-\ref{c0}) together with the NFW density profile,
without core, we have a model that can in principle be used to describe dark matter
structures with only one free parameter: $M_v$. A
particular case can always deviate from this average model mostly because
of the scatter in the concentration parameter $c$. This situation can
be alleviated if one takes into account the expected
range of values for $c$ given by $\sigma_{log c}\approx 0.095$ as described above.
 
Such description is, however, not valid for all dark
matter halos. There is specially a controversy, still not resolved, of whether or not
all galaxies are consistent with the cuspy behavior of the NFW profile in the center.
For dwarf and low surface brightness (LSB) galaxies there seems to be a disagreement
\cite{Blok1}-\cite{Bosma3}\nocite{Blok2,Bin,B-S,B-O}. On the other hand, for
large structures like galaxy clusters, the NFW description is strongly
supported by observations \cite{Pointe}-\cite{Zapa}\nocite{Voigt}. Recent numerical simulations  
are able to resolve regions closer to the center of dark matter halos and seem to 
favor even cuspier density profiles. For example in \cite{Nav04},
$\rho\sim r^{-\gamma}$ with $\gamma\sim$ 1.1, 1.2 and 1.35 for cluster-, galaxy- and dwarf-size 
halos respectively, another analysis (\cite{Die05}) finds $\gamma\sim1.2$ for a cluster-size halo.
Both analysis agree that a model with a core underestimates the simulated dark matter density
within the resolution limits of each work, this limit is roughly 1 percent of the
virial radius. The estimates for the parameter $\alpha$ that we obtain further on are still
consistent with these numerical results since the values for the core radius that we find for our
model are slightly inside this resolution limit.

For the purposes of our analysis, and given the hypothesis we have made, we
are interested in structures that are dominated by dark matter, that is,
that the baryonic or visible component is less representative and has no
important global effects. In fact, in order to empirically estimate the
entropy of dark matter halos, we have neglected the effects of the luminous
galaxies within them, thus, our model should
be in principle formally valid only for structures that are made exclusively of
dark matter. Dark matter dominated systems, like dwarf and LSB galaxies, whose dynamics is close to systems
which only have dark matter, will be taken as a reference to put bounds on 
the parameter $\alpha$. Taking into account the information we have on such systems, the choice of a halo model 
with the same properties of an NFW profile in the external region, but
with a core in the central region, seems to be reasonable.

We consider Eqs.~(\ref{delta0}-\ref{c0}) to be valid as well for this
model. Such assumption is strictly valid for Eqs.~(\ref{delta0}-\ref{rho0})
because they refer only to a particular way to define the normalization of the
profile, they can be used as long as one takes into account its connection
to the central density: $\delta_0\rho_0/(y_c(1+y_c)^2)=\rho_c$. The same reasoning is 
valid for Eq.~(\ref{r_v}), which is independent of the particular model for
the density profile, it is a definition for virialized spherical systems 
depending only on the cosmological model. Eq.~(\ref{c0}) is valid only
as an approximation, the virial mass appearing on it refers strictly to
the virial mass computed for an NFW model without core, the introduction of the
core reduces the mass in a percentage that depends on how large is $r_c$. 
However, for low values of $r_c$ compared to $r_v$, the difference caused by the
core is not relevant and the formula can be taken as a good approximation. In
any case, Eq.~(\ref{c0}) can be always considered as a fair way to give a
value to one of the free parameters ($c$ or $r_s$) in the model.

With all the previous considerations,
we can continue and calculate analytically the parameter $\alpha$. First, the
mass profile for our model follows from Eq.~(\ref{rho}):
\begin{displaymath}
\textrm{$M(y)=
\left\{\begin{array}{ll}
\frac{4}{3}\pi\rho_cr_s^3y^3 & \textrm{if $y < y_c$}\\
\frac{r_s}{G}V_0^2\left(\frac{y_c(4y_c+3)}{3(1+y_c)^2}+
ln\left(\frac{1+y}{1+y_c}\right)-\frac{y}{1+y}\right) & \textrm{if $y_c \leq y \leq y_v$}\\
\end{array} \right.$}
\end{displaymath}
\begin{equation}
\,\
\end{equation}
where $V_0^2=4\pi Gr_s^2\delta_0\rho_0$ and $G=4.297\times10^{-6}(\textrm{km/s})^2\textrm{kpc}/M_{\odot}$ 
is the gravitational constant in appropriate units (here we have followed 
closely the work \cite{Matos}). Using the mass profile, we calculate the gravitational 
potential at $r=0$, $\Phi(0)$: $d\Phi(r)/dr=GM(r)/r^2$; we use also the fact that $\Phi(r_v)=-GM(r_v)/r_v$
($\rho=0$ for $r>r_v$), we obtain:
\begin{equation}\label{phi0}
\Phi(0)=V_0^2\left(\frac{1}{1+y_v}-\frac{3y_c+2}{2(1+y_c)^2}\right).
\end{equation}
In order to calculate the velocity dispersion in the center, $\sigma(0)$, we
use the Jeans equation for spherical systems \cite{BT,Lokas}:
\begin{equation}\label{jeans}
\frac{1}{\rho}\frac{d(\rho\bar{v_r^2})}{dr}+2\frac{B\bar{v_r^2}}{r}
=-\frac{d\Phi}{dr},
\end{equation}
where $\bar{v_r^2}$ is the mean radial square velocity and 
$B=1-\bar{v_{\theta}^2}/\bar{v_r^2}$, with $\bar{v_{\theta}^2}$ the
mean square velocity in the $\hat\theta$ direction. $B$ is a measure of the
anisotropy in the system. It is worth saying that several studies have shown that dark
matter structures are actually anisotropic (see for instance \cite{Dehnen,Moore}); however, 
in the center of these structures, the studies usually give
$B\sim0$. We stay within the isotropic model for simplicity
considering that it will be a good approximation in the central regions
to real structures with a density core. Such
assumption implies also that $\bar{v_r^2}=\sigma^2$, where $\sigma$ is
the velocity dispersion, related to the temperature of
the neutralino gas. We solve then the Jeans equation for the two regions in 
which our profile is divided ($y<y_c$ and $y\geq y_c$) and match both solutions 
(recalling that $\rho(r_v)=\sigma(r_v)=0$). Thus,
the velocity dispersion in the center is: 
\begin{eqnarray}\label{sigma0}
&&\textrm{$\sigma^2(0) = \frac{V_0^2y_c(1+y_c)^2}{2}\left\{6(Li_2(1+y_v)-Li_2(1+y_c))+\right. $}\nonumber \\
&&\textrm{$ \textrm{ln}\left(\frac{1+y_v}{1+y_c}\right)\left[3\textrm{ln}\left(\frac{1+y_v}{1+y_c}\right)+
\frac{2y_c(4y_c+3)}{(1+y_c)^2}+\frac{y_v^3-5y_v^2-3y_v+1}{y_v^2(1+y_v)}
\right]$}\nonumber\\
&&\textrm{$\left. -\textrm{ln}\left(\frac{y_v}{y_c}\right)\left[6\textrm{ln}\left(\frac{1}{1+y_c}\right)
+1+\frac{2y_c(4y_c+3)}{(1+y_c)^2}\right]+\frac{1}{3}\left[\frac{1}{y_v(1+y_v)}\right.\right.$}\nonumber\\
&&\textrm{$\left. \left(\frac{y_c}{y_v(1+y_c)}
\left(-23y_v^3-41y_v^2-9y_v+3+\frac{y_c}{1+y_c}\left(-12y_v^3-18y_v^2
\right.\right.\right.\right.$}\nonumber\\
&&\textrm{$\left.\left.\left.\left.\left.-3y_v+1\right)+\frac{34y_v^3+47y_v^2+7y_v-3}{1+y_v}\right)\right)
\right]+
\frac{y_c^3(y_c+2)}{(1+y_c)^4}\right\}$}
\end{eqnarray}
where the dilogarithmic function is defined as: $Li_2(x)=\int_1^x\frac{\textrm{ln}~t}{1-t}dt$.

\subsection{Analysis on LSB galaxies}

Using Eqs.~(\ref{phi0}) and (\ref{sigma0}) we have an analytical expression for
the parameter $\alpha$. In order to give an empirical estimate for the range of values
that $\alpha$ can take, it is necessary to compute $V_0^2$, $y_c$ and $y_v$, which is
equivalent to give values for $r_v$, $r_s$ and $r_c$. To do so we use 
observational data extracted from the recent sample of LSB galaxies presented 
by \cite{Kuzio06}. Although LSB galaxies are 
dark matter dominated systems, baryons have non-negligible effects
in the total mass distribution of the galactic system, specially in the central 
region. A more realistic approach to these systems consists then in taking into
account the dynamical effects of baryons as an extra gravitational component, 
which changes the original dark matter distribution, pulling it inwards towards the center
during the process of disk formation (see for example \cite{Mo} for a description of 
a classical model that incorporates this effect). Such analysis is more complicated
and we consider that it would not increase significantly the accuracy of our model; 
the main reason is that the formula for the entropy of halos (Eq.~(\ref{SHALO})) was postulated
under the assumption of a single dark matter fluid, the addition of a baryonic component
could invalidate it, the mass distribution model should then be changed only if the 
entropy formula changes accordingly. We believe then that taking the so called ``minimum-disk'' 
approximation (or ``zero-disk''), that is ignoring the contribution of stars and gas, is 
consistent with the hypothesis we have used so far. As we emphasize later on, the 
values of the parameter $\alpha$ found using this approximation on the sample of LSB galaxies,
are roughly consistent with a particular analysis made for a more clear prototype of a dark
matter dominated galaxy, where the contribution of baryons can be safely ignored.

The sample presented in \cite{Kuzio06} consists of 11 LSB galaxies. We use their table 1 
to fit our halo model so that it has the same central density in the core, $\rho_c$ (computed by the
authors using a pseudoisothermal halo model), and finding the best fit to the whole rotation curve
of each galaxy in the sample. Following the model we have proposed, we find a unique fit to each 
galaxy for each value of the concentration parameter. To complete the analysis, we take the expected 
dispersion on the concentration value ($\sigma_{log c}\approx 0.095$) and make fits to the sample of 
galaxies using the extreme values of concentration given by this interval.

As a result of this analysis, we find a dynamical range in virial masses for the halo fits of:
$4\times 10^{10}M_{\odot} \leq M_v \leq 3\times 10^{13}M_{\odot}$, or
equivalently: $89$ kpc $\leq r_v \leq 810$ kpc; the range of values for
the concentration parameter goes from $6.6$ to $19.4$, and finally, for the
core radius, we find $0.3$ kpc $\leq r_c \leq 4.3$ kpc. If we divide the fitted halos
in three virial mass bins then the following intervals for $\alpha$ are found: 
$15.5\leq\alpha\leq32.9$, $12.7\leq\alpha\leq106.4$
and $17.8\leq\alpha\leq94.9$ for the low-, mid- and high-mass halos respectively. Since we would like to have an interval
for $\alpha$ which is scale independent, we take the final bounds for $\alpha$ to be:
\begin{equation}\label{alpha}
  17.8\leq\alpha\leq32.9
\end{equation}
Of the 11 galaxies in the sample, 7 can be fitted very well, 3 reasonably well and only 
1 poorly with values within this range. In a preliminary analysis to obtain the final interval for
$\alpha$, another sample of galaxies was also analyzed, that sample consisted of
6 dwarf, 9 LSB and 2 low-luminosity galaxies compiled in \cite{Firmani}. For that sample we used
a slightly different method to obtain $\alpha$, the two main differences are:
i) a different formula for the central value
of the concentration and its scatter (see \cite{Zavala}) and ii) instead of fitting the whole rotation
curve, only the value of the maximum rotational velocity was chosen to be the same; the interval for
$\alpha$ was found to be [16.4,27.8], which is similar to the one in Eq.~(\ref{alpha}). Such similarity
is found also with the work that motivated the present one \cite{lyr},
the authors used a different halo model and followed a completely different path to
obtain the range of values of $\alpha$, they found: $11.2\leq\alpha\leq24.8$. As
can be seen, despite the difference in the halo models and in the methods to obtain $\alpha$,
all these analysis end up with similar intervals for its value.

We consider that the method we have described so far is well suited to give an empirical
interval for $\alpha$, which is at the end one of the key parameters in our work. It is however
important to say that $\alpha$ is in principle a model dependent quantity whose
value could change by assuming other halo models, or by incorporating other effects, for example the addition of
the baryonic component or the inclusion of anisotropy. As have been said before, the formula for the entropy of halos 
(Eq.~(\ref{SHALO})), which is the basis for the final constraints we have found on the mSUGRA parameter space
(see next section), 
is strictly valid only for dark matter halos without baryons. So a better alternative to get closer to this condition
is to analyze dwarf spheroidal (dSph) galaxies, which are dark matter dominated systems with very high 
mass-to-light ratios (higher than a 100 $M_\odot/L_\odot$). To extend our analysis further and be more 
certain on the numerical values of $\alpha$, we analyze a specific case, the Draco dSph galaxy (hereafter, Draco).
  
\subsection{A particular case analysis for dSph galaxies: Draco}

\begin{figure*}\centering
  \includegraphics[height=10cm, width=12cm]{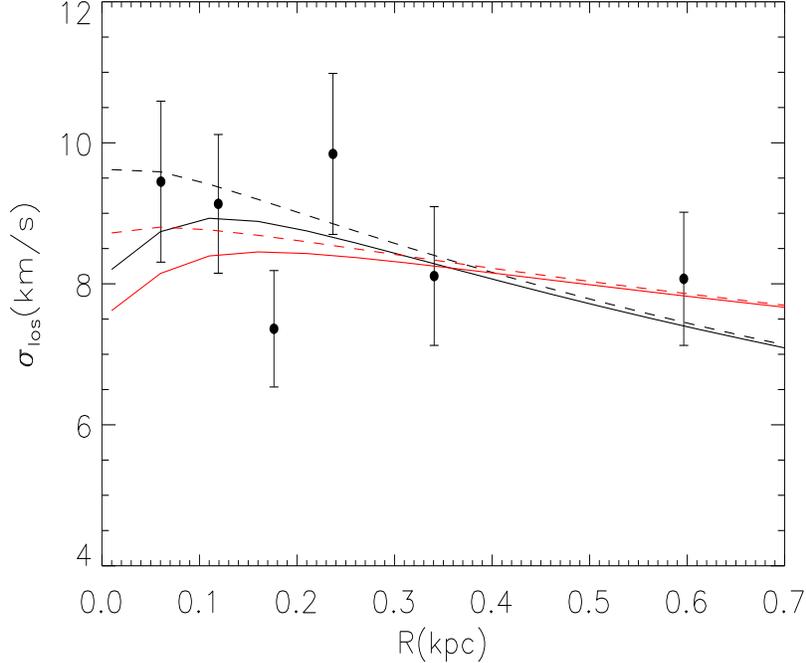}
  \caption{The line-of-sight star velocity dispersion as a function of the projected radius for Draco.
    Observations were taken from \cite{Prada}. The solid and dotted lines are fits from our halo model, see text
    for details.}
  \label{sig_draco}
\end{figure*}

Draco is a good generic example of dSph galaxies since it has been the subject of extensive 
observational studies (see \cite{Wilki} for a recent one). The observational quantity measured for these
type of systems is the line-of-sight velocity dispersion of its stars, $\sigma_{los}$, as a function of their
projected radius. We use the observational values reported in \cite{Prada} for $\sigma_{los}$ (extracted
from the upper panel of their fig. 5). They took the original observational stars sample of \cite{Wilki} and 
removed unbound stars following a rigorous method, see \cite{Prada} for details. The observational data
appears on figure \ref{sig_draco}.

To compute $\sigma_{los}$ for our halo model, we follow the formulas described
in \cite{Lokas_sig1} and \cite{Lokas_sig2}. Of most importance is the use of the integral formula for
the line-of-sight velocity dispersion presented in the appendix of \cite{Lokas_sig2} (isotropic case, $B=0$):
\begin{equation}\label{sigma_los}
  \sigma^2_{los}(R)=\frac{\Gamma(2m)}{\Gamma((3-p)m)}\frac{r_v}{a_s}V_v^2\textrm{exp}(X^{1/m})
  \int_X^\infty\sqrt{1-\left(\frac{X}{x}\right)^2}\tilde{\ell}(x)\tilde{M}(x)\frac{dx}{x}
\end{equation}
where $\tilde{\ell}(x)=x^{-p}\textrm{exp}(-x^{1/m})$ is the dimensionless 
3D luminosity density of the stellar component 
(which can be obtained by deprojecting the surface brightness profile, modeled in \cite{Lokas_sig1} 
by a S\'ersic law), $p=1-0.6097/m+0.05463/m^2$; 
$\tilde{M}(x)=\tilde{M}_{DM}(x)+f_\star\tilde{M}_{\star}(x)$ is the dimensionless mass profile consisting 
of dark matter, $\tilde{M}_{DM}$, and stellar, $\tilde{M}_{\star}$, components; $a_s$ and $m$ are 
the free parameters of the S\'ersic profile, and finally $x=r/a_s$, $X=R/a_s$. Eq. (\ref{sigma_los})
is generically valid for any halo model with spherical symmetry. The stellar component has 3 parameters:
$a_s$, $m$ and $f_\star$, the first two are constrained by the brightness profile of Draco \cite{Ode}: 
$a_s=7.3$ arcmin ($\sim0.17$ kpc for a distance to Draco of 80 kpc) and $m=0.83$. And $f_\star=M_\star/M_v$,
where $M_\star=6.6\times10^5M_\odot$ (\cite{Lokas_sig1}). 

Using Eq.~(\ref{sigma_los}) together with our halo model, we can find fits to the observational data using
2 free parameters: $M_v$ and $\rho_c$, and using the central value for the concentration parameter 
(Eq.~(\ref{c0})) altogether with its expected dispersion $\sigma_{log c}\approx 0.095$. In Fig.
(\ref{sig_draco}) we show different fits to the observational data, the black (thicker) and red (thinner) lines are for
the upper and lower values of the concentration parameter respectively. We found that for the lower
values of the concentration parameter, red (thinner) lines, the best fit to the data is a model which converges
towards a halo with no core (red (thinner) dashed line). Increasing the core radius reduces the goodness of the 
fit, we measured this evaluating $\chi^2/df$, where $df$ is the number of degrees of freedom: number
of data points - number of parameters. On the contrary, for higher values of the concentration parameter, 
black (thicker) lines, the best fit is found for a model with core (black (thicker) solid line), the goodness of the fit
is statistically reduced with decreasing core radii, the convergence to a halo with no core is shown
as a dashed black (thicker) line. These results imply that our model with high values of the concentration 
parameter is able to give an adequate description of the actual observations for Draco.

Our whole model is built on the assumption of a model with core,
this goes back to Eqs. (\ref{shalo}) and (\ref{SHALO}), so we should
clearly concentrate on the fits corresponding to a model with a
core. In the search for the best fit to $\sigma_{los}$, this can be
done unambiguously only for certain values of the concentration
(those on the high end of the expected concentration interval). For
instance, the red (thinner) solid line is as good fit to the
observational data as the black (thicker) solid line, however, we
can not extract the value of $\alpha$ without ambiguity for the
latter since reducing the core radius increases the goodness of the
fit, recall that convergence is reached for $r_c\rightarrow0$ kpc.
The core model which we find to be a good fit to the observational
data (black (thicker) solid line) has the following parameters:
$M_v=1.5\times10^9M_\odot$, $\rho_c=3.1\times10^8M_\odot$kpc$^3$,
$r_c=0.11$kpc, $r_v=30$kpc, $c=26.6$ and $\alpha=28.7$. 

We believe
that this value of $\alpha$ for Draco is reasonable and certainly
consistent with the interval obtained before (Eq.~(\ref{alpha})). 
As a complementary check, we have used another set of data on the velocity dispersion
of Draco (\cite{Walker07}) to fit our model. Although larger values for the virial mass
and core radius are needed to achieve a good fit, we found the value of $\alpha$ to be consistent
with the interval of values given in (Eq.~(\ref{alpha}).
Nevertheless it should be stressed that better observational
constraints are needed to put firmer constraints on the value of
$\alpha$, which is particularly sensitive to the value of $\rho_c$
($r_c$). The value of $\alpha$ is of course also sensitive to the
chain of assumptions needed to arrive to its final estimate,
concentration values have an important role as can be seen on figure
\ref{sig_draco}. The assumption of isotropic velocity dispersion, B=0 in Eq.~(\ref{jeans}), 
has also an influence in the values found for the model parameters in the fit to Draco. The addition
of anisotropy to our model would alter the values of the best fit parameters found for Draco, and 
therefore the value of $\alpha$ we just reported. The impact of anisotropy on the value of $\alpha$
can only be properly quantified by improving our model with the removal of the restriction
B=0, and redoing the analysis we have made so far. Such task   
is left as a possible future work once further more precise constraints on the 
dynamical properties of dark matter dominated systems like Draco are available. We believe that 
the assumption of isotropy is sufficient for the purposes of this work to calculate approximately 
the value of $\alpha$.
 
Overall, taking into account all the assumptions we
have made to fit our model with observational data, we consider that
the interval of values obtained for the ``ignorance parameter''
$\alpha$, do give a reasonably description to the several galactic
samples for which we have tested our model, thus making us feel
confident on the overall entropy criteria presented in this work, at
least as a first order indirect analysis of real dark matter
structures.

\section{Analysis of the mSUGRA parameter space}

Supersymmetry has attracted attention since it was first proposed
for its many intriguing theoretical features and also 
for providing an elegant solution to the hierarchy problem (for
an excellent introduction to the supersymmetry formalism and 
phenomenology see \cite{martin}).

The minimal phenomenologically viable supersymmetric extension of the
SM is called the Minimal Supersymmetric Standard Model (MSSM).  The
MSSM has appealing features: it is in much better agreement with the
assumption of unification of the gauge couplings than the Standard
Model \cite{Amaldi-Kim}, besides, it provides with a good dark
matter candidate when the lightest supersymmetric particle (LSP) is
the neutralino (see for instance \cite{Jungman:1995df}).  In the MSSM
every known particle is associated to a superpartner to form either a
chiral or a gauge supermultiplet, whose spin components differ from
each other by $1/2$.  In addition it has two $SU(2)$ doublet complex
scalar Higgs fields ($H_u$ and $H_d$) in order to give masses to the
up and down type particles and to avoid gauge anomalies.  The ratio of
the vacuum expectation values of the neutral part of these two Higgs
doublets is known as $\tan\beta = <H^0_u>/<H^0_d>=v_u/v_d$.  The MSSM
includes a discrete matter parity called R parity.  All the SM
particles have charge $+1$ under this symmetry, whereas the
supersymmetric partners have charge $-1$. R parity is responsible for
the stability of the LSP.

Since we do not observe any of the
superpartners at the current laboratory energies, if supersymmetry
exists it has to be broken.  The breaking of SUSY introduces around
$\sim 100$ new parameters, called soft breaking SUSY terms.  These are
constrained by the limits on flavour changing neutral currents and
CP-violating processes, which lead to the assumption of
``universality'' of the soft breaking terms.  The universality
condition means that the gaugino masses have a common value at the GUT
scale, and the same assumption is made for the scalar (squark and
sleptons) masses and the trilinear couplings respectively. 
After SUSY and electroweak symmetry breaking, there is mixing
between different squarks, sleptons, and Higgses with the same
electric charge, and between the gauginos and electroweak Higgsinos.
Thus, the low energy mass spectrum of the MSSM contains the usual SM
particles, squarks and sleptons, four neutralinos (denoted by
$\chi_{1,2,3,4}^0$), two charginos ($\chi^{\pm}_{1,2}$), the gluino,
and five Higgs bosons: the usual light SM-like Higgs $M_{Higgs}$, a 
heavy neutral one $M_H$, two charged ones $M_H^{\pm}$, and a pseudoscalar $M_A$.

Minimal supergravity (mSUGRA) 
is one of the better studied SUSY breaking scenarios
\cite{Nilles:1983ge}.  In this class of models supersymmetry is broken
spontaneously in a hidden sector that connects only through
gravitational-strength interactions with the MSSM or visible sector.
In the visible sector these interactions induce the appearance of the
soft SUSY breaking terms. These are determined by only five
parameters, a universal mass for the gauginos at the GUT scale
$m_{1/2}$, a universal mass for the scalars $m_0$, a universal
trilinear coupling $A_0$, $\tan\beta$, and the sign of the Higgsino mass
parameter $\mu$.  This drastic reduction in the number of parameters
facilitates the scanning of interesting regions of the parameter
space.  This is usually done by fixing two parameters, for instance
$A_0$ and $\tan\beta$. 
It is also possible to vary the four continuous
parameters freely using Monte Carlo techniques with interesting
results \cite{Baltz:2004aw,Allanach:2005kz}.

We are finally in a position to use both the abundance and entropy criteria, 
to compute the total mass density of neutralinos today,
and constrain the region in the mSUGRA parameter space where
both criteria are fulfilled. Out of the five parameters of the model, we will
consider that the Higgsino mass parameter has a positive sign. This
consideration is based on studies of the anomalous magnetic moment of the muon
$g_\mu -2$, where SUSY models with a positive sign of $\mu$
can give a much better agreement with the experimental value of $g_\mu
-2$ than in the Standard Model, whereas negative values do not solve
this problem \cite{g-2exp2}-\cite{g-2CNH}\nocite{DDDD,g-2MSSMf1l}.

Our strategy is then to explore broad regions of values for 
the other four parameters, by means of a bidimensional analysis
in the $m_0 - m_{1/2}$ plane with different fixed values
of $A_0$ and tan$\beta$. It is important to mention that we are not
presenting an exhaustive search in all the possible regions, but
we concentrate on those regions which have received more attention
in the literature, see for example \cite{belanger}.

\begin{figure*}\centering
\includegraphics[height=16cm, width=4.8cm, viewport=200 0 300 350]{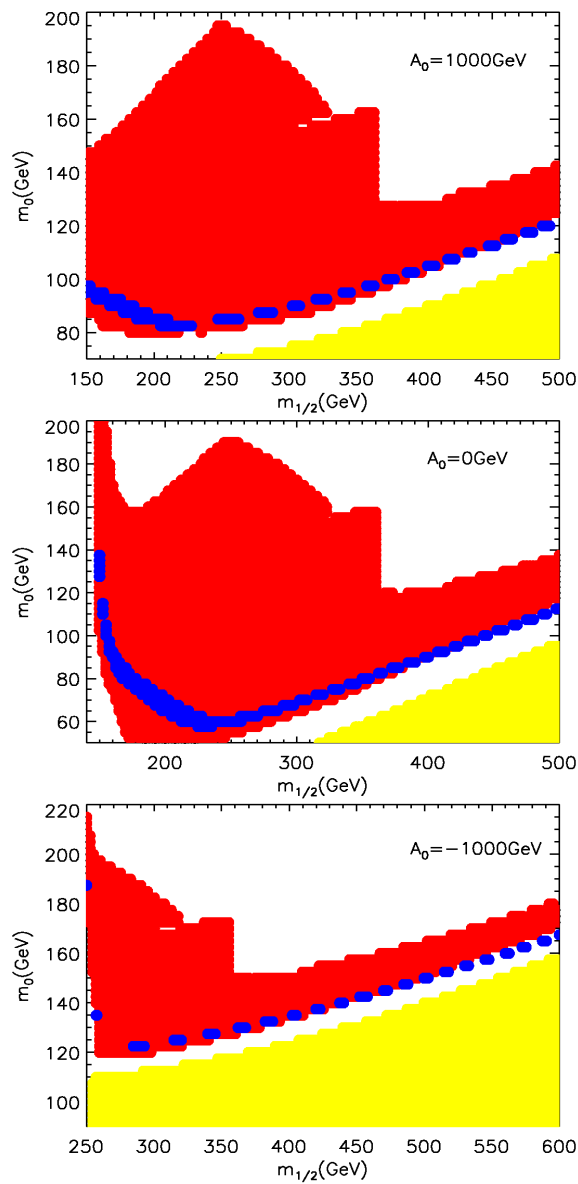}
\caption{Allowed regions in the parameter space for AC (lighter gray, red) and EC (darker grey, blue) for 
the mSUGRA model with sgn$\mu=+$, tan$\beta=10$, and $A_0=1000$ GeV, top panel, $A_0=0$ GeV, middle panel, 
and $A_0=-1000$ GeV, bottom panel. The yellow region (lower right corner) is where the $\tilde{\tau}$ 
is the LSP. The figure shows the so called bulk and coannihilation regions.}
\label{tb10new}
\end{figure*} 

In Fig.~(\ref{tb10new}), we present the results for tan$\beta=10$, and for three
values of $A_0$, namely $A_0=1000, 0, -1000$ GeV, shown in the top, middle and
bottom panels respectively. The figure shows the so called bulk and coannihilation regions.
The yellow region (lower right corner) is where the stau $\tilde{\tau}$ is the LSP, the lighter and darker areas 
(red and blue for the online version in colors) define the allowed regions for the EC and AC
respectively according to the observed value of $\Omega_{DM}$. The
area of the EC region depends on the size of the interval of values
of the parameter $\alpha$, Eq.(\ref{alpha}), the lower and upper bounds of
$\alpha$ determine the upper and lower boundaries of the EC region. As can be seen from the figure,
the region where both criteria are fulfilled is very small,
in fact, only for the highest values of $\alpha$ there is an
intersection between both criteria. This behavior holds for all values of $A_0$ in the interval
$[-1000, 1000]$ GeV; here we are showing the results only for the extreme and central values of 
$A_0$\footnote{For Fig. (\ref{tb10new}) and the following Fig. (\ref{results2}), the disconnected
regions are caused by discreteness on the grid values chosen to explore the $m_0-m_{1/2}$ plane.}.

\begin{figure*}\centering
\includegraphics[height=16cm, width=4.8cm, viewport=200 0 300 350]{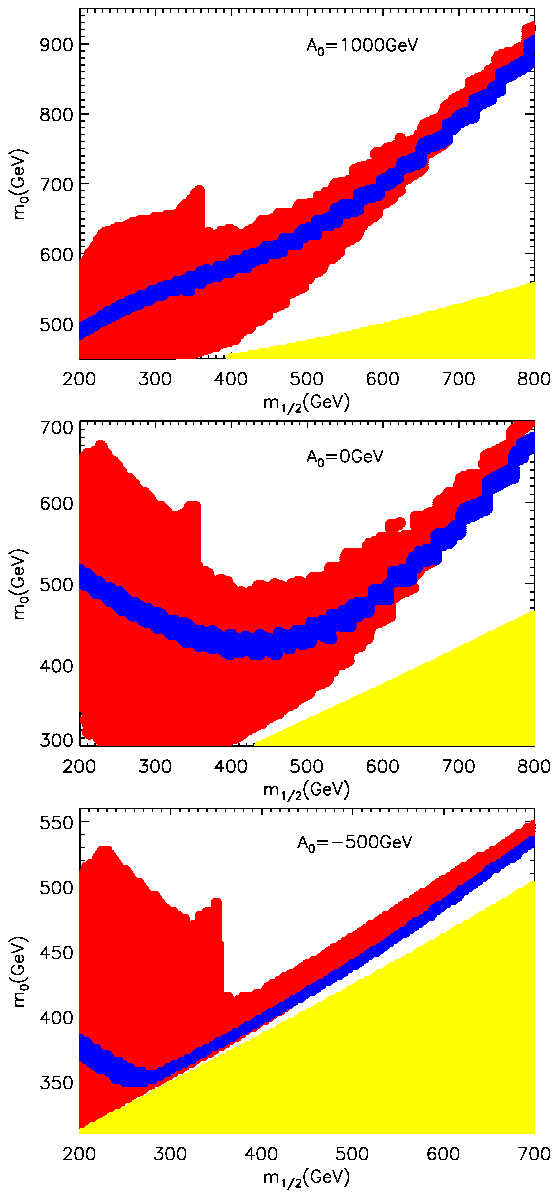}
\caption{The same as Fig.~(\ref{tb10new}), but for tan$\beta=50$, and now $A_0=-500$ GeV in the
bottom panel.}
\label{tb50new}
\end{figure*} 

Repeating the same procedure for larger values of tan$\beta$, we find
that the intersection region for both criteria becomes larger, getting
more significant for larger values of this parameter. This is clearly
shown in Fig.~(\ref{tb50new}), equivalent to Fig.~(\ref{tb10new}), but
for tan$\beta=50$. In this case the bottom panel is for $A_0=-500$
GeV. It is clear from the figure that for this value of tan$\beta$,
both criteria are consistent, there is a large intersection area for
values of $A_0$ in the interval $[0,1000]$ GeV. For negative values of
$A_0$, the intersection region decreases as $A_0$ does so, see the
bottom panel of Fig. (\ref{tb50new}). For even lower values of $A_0$
the intersection becomes insignificant.

\begin{figure*}\centering
\includegraphics[height=14cm, width=16cm]{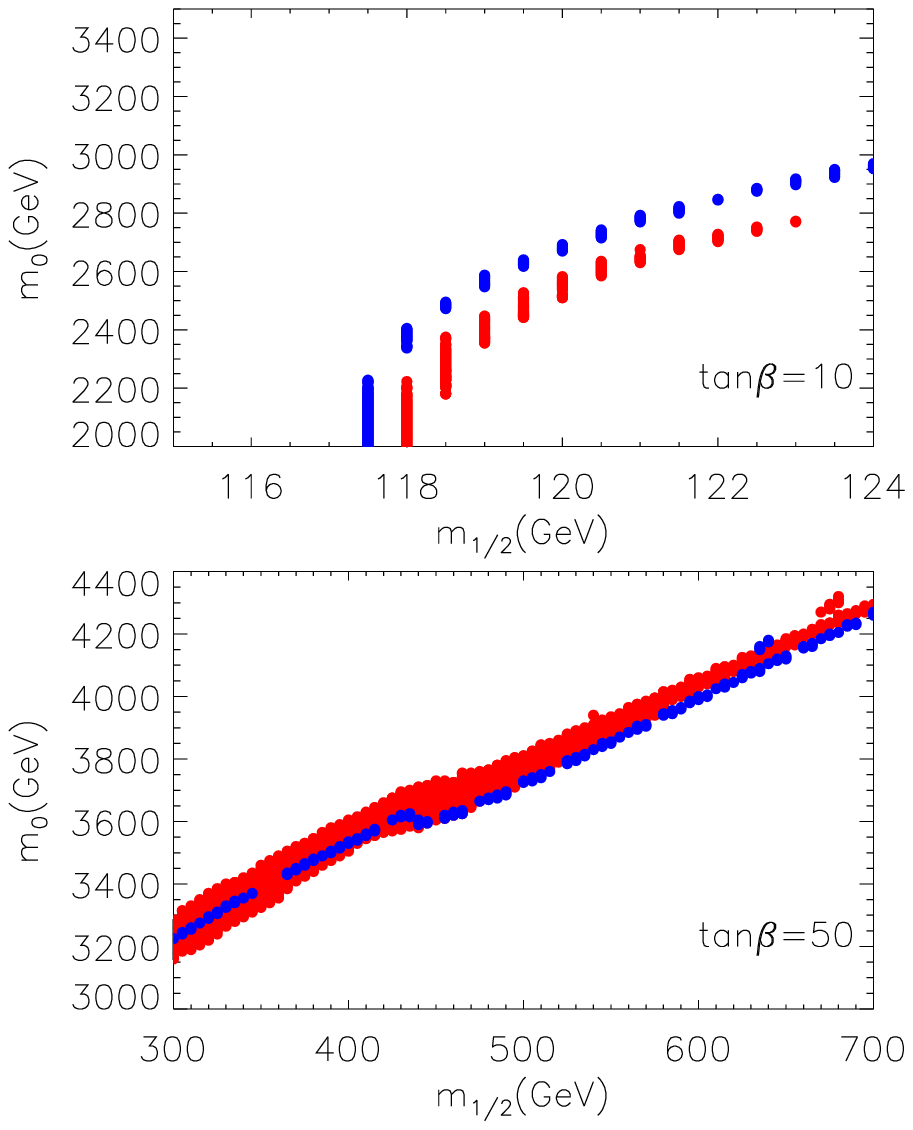}
\caption{Allowed regions in the parameter 
space for AC (lighter gray, red) and EC (darker grey, blue) in the 
mSUGRA model with $A_0=0$, sgn$\mu=+$, tan$\beta=10$, top panel, and tan$\beta=50$, bottom panel. 
The figure shows the so called Focus Point region.}
\label{results2}
\end{figure*}

In Fig.~(\ref{results2}) we present the same analysis but for a
different region of the parameter space (high values of $m_0$), known
as Focus Point region, and for the central value $A_0=0$.  The
situation is consistent with the previous result, both criteria
intersect for $\tan \beta=50$ and there is no intersection for
$\tan \beta=10$.

This analysis allows us to arrive to one of the main results of our
work. The use of both criteria favors large values of tan$\beta$.

It is interesting to point out that, within the AC formulation, the
dark matter density constraint reduces effectively the allowed regions
in the mSUGRA parameter space to strips pointing to an almost linear
trend between $m_0$ and $m_{1/2}$, at least in the coannihilation and
focus point regions, as can be clearly seen in the darker (blue) areas
in Figs.~(\ref{tb10new}, \ref{tb50new}, \ref{results2}).  This type of
behavior has been noticed previously in \cite{Batt}, where the authors
even give some parameterizations for these so called ``WMAP lines''.
Notice that this type of relation is also shown within the EC
description in the above mentioned regions, although the ``WMAP
lines'' are not as narrow as in the AC formulation, because of the
wide range in the $\alpha$ parameter Eq.~(\ref{alpha}). A deeper
analysis of the connections among the mSUGRA parameters could reveal
the original cause for such behavior, shown in both, quite
independent, criteria.

\begin{figure*}[htb!]
\centerline{\includegraphics[width=8cm,angle=0]{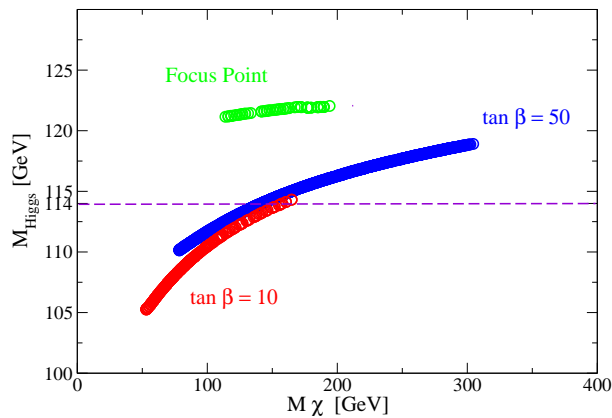}}
\caption{The figure shows the Higgs mass plotted versus the LSP mass.
  Points above the dashed line are the allowed values for the LSP.}
\label{fig:chi-higgs1}
\end{figure*}

The regions that were compatible with the abundance and entropy
criteria were analyzed to see what restrictions they imposed on the
SUSY spectra. This was done just in order to show the expected mass spectra
resulting from these constraints. The analysis was performed using micrOMEGAs linked to
Suspect \cite{suspect}, using the default input values of micrOMEGAs, for instance:
\begin{eqnarray}
M_{\rm top} &=& 175 \, {\rm GeV},\\
m_{\rm bot} &=& 4.23 \, {\rm GeV}.
\end{eqnarray}
The Higgs bosons masses, as well as the rest of the superpartner
masses, depend directly on the Universal gaugino mass $m_{1/2}$ and
the Universal scalar mass $m_0$.  The first noticeable fact is that
the bound on the Higgs mass \cite{Higgs}:
\begin{equation}
m_{\rm Higgs} \geq 114 \,{\rm GeV,}
\end{equation}
favors the results with large $\tb$ for the bulk and co-annihilation
regions, as can be seen from Fig.~(\ref{fig:chi-higgs1}), which shows
the Higgs mass $M_{Higgs}$ plotted against the mass of the LSP.  
In the case of both the bulk and co-annihilation
regions, for   $\tb = 10$ only a small region of
parameter space is allowed, with $m_{\chi} \sim 160$ GeV.  In this same case
for $\tb =50$ the allowed SUSY spectra starts for $m_{\chi} \geq 141$
GeV.  In the case of the focus point region the Higgs mass does not
impose any further constraint, and $m_{\chi} \geq 114$ GeV.
\vspace{0.0cm}
\begin{figure*}[htb!]
\centerline{\includegraphics[height=18cm, width=8cm]{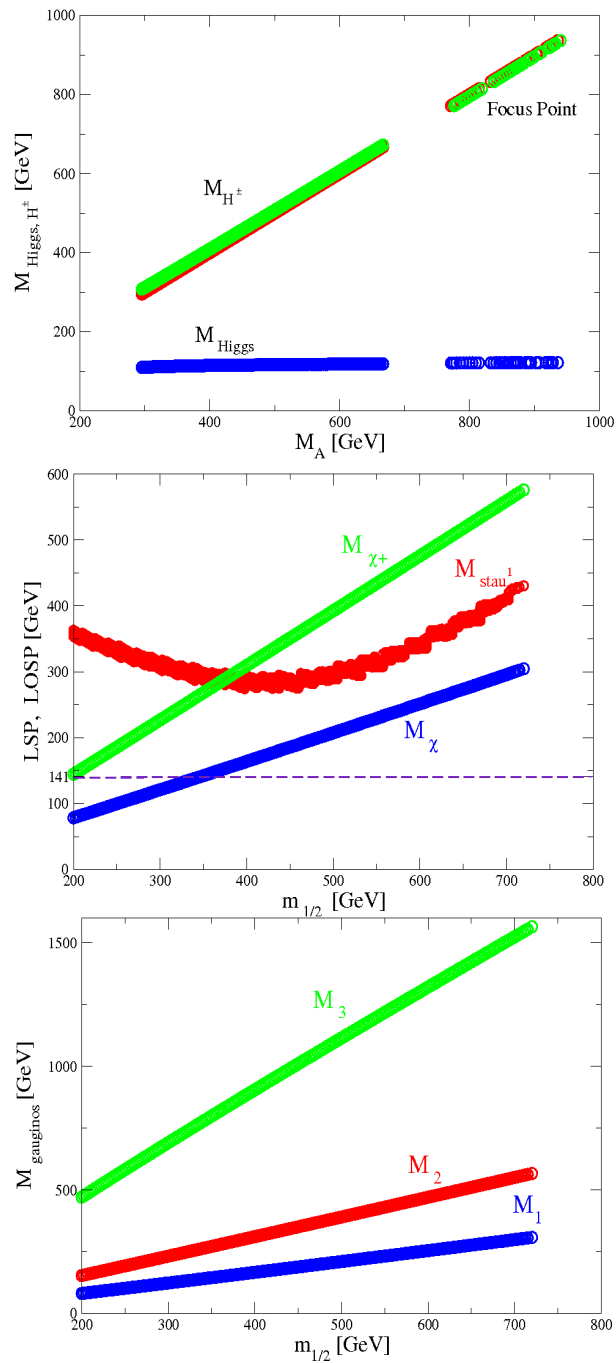}}
\caption{The lightest $M_{Higgs}$  and heavier Higgs bosons
$M_H,~M_H^{+}$  plotted against  $M_A$ for $\tb=50$ (top panel). The  lines to the right of the graph correspond to
the focus point region, where the masses are in general heavier
(except for $M_{Higgs}$). The mass of the LOSP plotted against $m_{1/2}$ for
$\tb=50$ in the bulk and
co-annihilation regions (middle panel). The masses of the gauginos $M_1,~M_2,~M_3$ 
plotted against $m_{1/2}$ for $\tb=50$ also in the bulk and
co-annihilation regions (bottom panel).}
\label{fig:mAvsmhs}
\end{figure*}

Fig.~(\ref{fig:mAvsmhs}) top panel, shows the different Higgs bosons
masses plotted against the pseudoscalar one $M_A$, for $\tb=50$.  As
can be seen from the graph, the heavy neutral $M_H$ and charged
$M_{H^{+}}$ Higgs bosons are much heavier than the lightest one, and
almost degenerate in mass.  The lightest supersymmetric observable
particle LOSP is plotted against $m_{1/2}$ for $\tb=50$ for the bulk
and co-annihilation regions. The region above the dashed line shows the
allowed values after the constraint on $M_{Higgs}$ is taken into
account. A comparison of the LSP with the LOSP shows that for this
region, for smaller values of $m_{1/2}$ the LOSP will be the lightest
chargino $\chi^\pm$ (which is practically degenerate in mass to the
second lightest neutralino), and very close in mass with the stau. As
$m_{1/2}$ increases the LOSP is the stau, again with values very close
to the LSP, as expected.  For the focus point region the LOSP is the
lightest chargino $\chi_1^{\pm}$, followed by the second neutralino
$\chi^0_2$, with almost degenerate masses.  Similar results hold for
$\tb=10$ albeit for a much more reduced region in parameter space.
The gauginos $M_1,~M_2,~M_3$ are plotted against $m_{1/2}$ in the
bottom panel for $\tb=50$. The masses of these particles on the focus
point region are very similar to the ones shown in the figure (they
basically lie along the same lines).  In both, large and small
$\tb$ cases, the gluino is much heavier than the other two gauginos,
as expected, and this difference increases with a larger value of the
neutralino mass, or equivalently, with larger $m_{1/2}$.

\section{Conclusions}

We have followed the novel idea of \cite{lyr} to introduce a new
criterion to constrain the mSUGRA parameter space, which uses the assumption
of entropy consistency for the initial and final states of a
neutralino gas. Using the program micrOMEGAs, we explored with
precision which regions simultaneously satisfy this criterion and the usual abundance
criterion previously used in the literature. 

Combining both, and using one of the most recent analysis that combines
observations coming from different sources to constrain the actual 
dark matter density, we were able to show that both criteria are compatible
for large values of $\tan \beta$, in particular for tan$\beta=50$, and that
for lower values ($\tan \beta=10$), the coincidence is scarce. The 
result then favors the scenario where the ratio of the vacuum expectation 
values for the neutral supersymmetric Higgs bosons is large. 
Some other SUSY extensions of the SM, taken together with LEP data \cite{:2001xx},  can also 
favor large values of $\tan\beta$ (for instance Grand Unified or supergravity  models 
with Yukawa coupling unification (see for instance \cite{Olechowski:1988gh, Ananthanarayan:1992cd}), or Finite 
Unified Theories (see for instance \cite{Djouadi:2004ae, Kapetanakis:1992vx})).

Moreover, we went a step further and analyze the mass spectrum of the
SUSY particles and obtained a bound for the neutralino mass for the
bulk and co-annihilation region, $m_\chi
\geq 141$ GeV, which is higher than the actual experimental
constraint, $m_\chi > 46$ Gev \cite{Yao:2006px}.

In the process to obtain the presented results, we described carefully a method to obtain an empirical
interval on the values for $\alpha$, a parameter reflecting the present ignorance on the appropriate statistical-mechanics
description on dark matter systems. We used a halo model with a constant density core in the center followed by an NFW
profile to describe dark matter dominated systems, which are the astrophysical systems for which our model is
suitable. We found that, although $\alpha$ is a model dependent quantity, the bounds reported in this paper for its
value (Eq.~(\ref{alpha})) are reliable, considering the different range of observational data we
used to constrain its value. More precise observations on the line-of-sight velocity dispersion of dwarf spheroidals
are promising to improve our results.

Finally, we want to remark that the analysis presented in this work is robust
and it can be done for any other particle claimed to be candidate for
dark matter, or for other supersymmetric extensions of the Standard Model.
Also, our results can be extended for other regions of the parameter space, for instance,
analyzing $A_0\neq0$ for the Focus Point region, $\mu<0$ and exploring the ``rapid annihilation funnel'' 
region (see for example right panel, Fig. (4) of \cite{Feng}). 

\ack 
We acknowledge partial support by CONACyT M\'exico, under grant 
U47209-F,  I0101/131/07 C-234/07 and grants PAPIIT-IN115207-2, and IN113907 by UNAM. DN acknowledges
partial support by DAAD and DGAPA-UNAM grants, and thanks the AEI-Max Planck in Potsdam for
warm hospitality during the ellaboration of this work. JZ acknowledges 
support from DGEP-UNAM and CONACyT scholarships, and support by the CAS Research Fellowship for 
International Young Researchers. JZ is supported by the 
Joint Program in Astrophysical 
Cosmology of the Max Planck Institute for Astrophysics and the Shanghai Astronomical Observatory.  
The numerical runs were performed in the clusters Tochtli of the Nuclear Sciences 
Institute at UNAM and Ekbek in LaSumA at CINVESTAV in Mexico City.


\pagebreak


\section*{References}

\begin{thebibliography}{90}

\bibitem{WMAP}Bennett C L et al. (The WMAP collaboration), {\it First-Year Wilkinson Microwave Anisotropy 
Probe (WMAP) Observations: Preliminary Maps and Basic Results}, 2003 {\it Astrophys. J. S. S.} {\bf 148} 1 
[astro-ph/0302207]\\
See also Spergel D N et al (The WMAP collaboration), {\it First-Year Wilkinson Microwave Anisotropy Probe 
(WMAP) Observations: Determination of Cosmological Parameters}, 2003 {\it Astrophys. J. S. S.} {\bf 148} 175 
[astro-ph/0302209];\\
Spergel D N et al. (The WMAP collaboration), {\it Three-Year Wilkinson Microwave Anisotropy Probe (WMAP) 
Observations: Implications for Cosmology}, 2007 {\it Astrophys. J. S. S.} {\bf 170} 377 
[astro-ph/0603449]
\bibitem{seljak}Seljak U, Slosar A and McDonald P, {\it Cosmological parameters from combining the 
Lyman-$\alpha$ forest with CMB, galaxy clustering and SN constraints}, 2006 {\it J. Cosm. Astropart. Phyis.} 
{\bf 10} 014 [astro-ph/0604335]
\bibitem{martin}
  M.~F.~Sohnius,
  Phys.\ Rept.\  {\bf 128} (1985) 39;
Martin S P,{\it A Supersymmetry Primer}, 1998 {\it Perspectives on Supersymmetry. Advanced 
Series on Directions in High Energy Physics. Edited by Gordon L. Kane. Published by World Scientific Publishing 
Company, Singapore} {\bf 18} 1 [hep-ph/9709356]
\bibitem{belanger}B\'elanger G, Kraml S and Pukhov A, {\it Comparison of supersymmetric spectrum 
calculations and impact on the relic density constraints from WMAP}, 2005 {\it Phys. Rev. D} 
{\bf 72} 015003 [hep-ph/0502079] 
\bibitem{constraint}B\'elanger G, Boudjema A, Pukhov A, and Semenov, A, {\it WMAP constraints on SUGRA models 
with non-universal gaugino masses and prospects for direct detection}, 2005 {\it Nucl. Phys. B} 
{\bf 706} 411 [hep-ph/0407218]
\bibitem{micro}B\'elanger G, Boudjema A, Pukhov A, and Semenov, A, {\it micrOMEGAs: Version 1.3}, 2006 
{\it Comput. Phys. Commun.} {\bf 174} 577 [hep-ph/0405253]
\bibitem{lyr}Cabral-Rosetti L G, Hern\'andez X and Sussman R A, {\it Using entropy to discriminate annihilation 
channels in neutralinos making up galactic halos}, 2004 {\it Phys. Rev. D} {\bf 69} 123006
\bibitem{proceeding}Nunez D, Sussman R A, Zavala J, Nellen L, Cabral-Rosetti L G and Mondragon, M, 
{\it Entropy considerations in constraining the mSUGRA parameter space}, 2006 {\it 10th Mexican Workshop 
on Particles and Fields, Morelia, Mexico, AIP Conference Proceedings 2005} {\bf 857} 321 [astro-ph/0604127]
\bibitem{kamion}Jungman G, Kamionkowski M and Griest K, {\it Supersymmetric dark matter}, 1996 
{\it Phys. Rept.} {\bf 267} 195 [hep-ph/9506380]
\bibitem{gondolo}Gondolo P and Gelmini G, {\it Cosmic abundances of stable particles: Improved analysis}, 
1991 {\it Nucl. Phys. B} {\bf 360} 145
\bibitem{suspect}Djouadi A, Kneur J L and Moultaka G, {\it SuSpect: A Fortran code for the Supersymmetric 
and Higgs particle spectrum in the MSSM}, 2007 {\it Comput. Phys. Commun.} {\bf 176} 426 [hep-ph/0211331]
\bibitem{Tsallis}Tsallis C, {\it Nonextensive statistics: Theoretical, experimental and 
computational evidences and connections}, 1999 {\it Braz J Phys} {\bf 29} 1 [cond-mat/9903356]
\bibitem{Tsallis2p}Nunez D, Sussman R A, Zavala J, Cabral-Rosetti L G and Matos T, {\it Empirical testing 
of Tsallis' Thermodynamics as a model for dark matter halos}, 2006 {\it 10th Mexican Workshop on Particles 
and Fields, Morelia, Mexico, AIP Conference Proceedings 2005} {\bf 857} 316 [astro-ph/0604126]
\bibitem{Tsallis2}Zavala J, Nunez D, Sussman R A., Cabral-Rosetti L G and Matos T, {\it Stellar polytropes 
and Navarro Frenk White halo models: comparison with observations}, 2006 {\it J. Cosm. Astropart. Phys.} 
{\bf 06} 008 [astro-ph/0605665]
\bibitem{Pathria}Pathria R K, {\it Statistical Mechanics}, 1972 (Pergamon Press)
\bibitem{BT}Binney J and Tremaine S, {\it Galactic Dynamics}, 1987 (Princeton University Press
\bibitem{Padma3}Padmanabhan T, {\it Theoretical Astrophysics, Volume I: Astrophysical Processes}, 2000
(Cambridge University Press)
\bibitem{Katz1}Katz J, Horwitz G and Dekel A, {\it Steepest descent technique and stellar equilibrium 
statistical mechanics. IV - Gravitating systems with an energy cutoff}, 1978 {\it Astrophys. J.} 
{\bf 223} 299
\bibitem{Katz2}Katz J, {\it Stability limits for 'isothermal' cores in globular clusters}, 1980 
{\it Mon. Not. R. Astron. Soc.} {\bf 190} 497
\bibitem{MPV}Magliocchetti M, Pugacco G and Vesperini E, {\it Gravothermal catastrophe in anisotropic 
systems}, 1997 {\it Nuovo Cimento B} {\bf 112} 423 [astro-ph/9806233]
\bibitem{Amaldi-Kim}
U.~Amaldi, W.~de Boer, and H.~Furstenau, 
  Phys.\ Lett.\ B {\bf 260}, 447 (1991);
C.~Giunti, C.W.~Kim and U.W.~Lee, 
  Mod.\ Phys.\ Lett.\ A {\bf 6}, 1745 (1991);
\bibitem{Jungman:1995df}
  G.~Jungman, M.~Kamionkowski and K.~Griest,
  Phys.\ Rept.\  {\bf 267} (1996) 195
  [arXiv:hep-ph/9506380].

\bibitem{Nilles:1983ge}
  H.~P.~Nilles,
  Phys.\ Rept.\  {\bf 110} (1984) 1.


\bibitem{Baltz:2004aw}
  E.~A.~Baltz and P.~Gondolo,
  JHEP {\bf 0410} (2004) 052
  [arXiv:hep-ph/0407039].

\bibitem{Allanach:2005kz}
  B.~C.~Allanach and C.~G.~Lester,
  Phys.\ Rev.\  D {\bf 73} (2006) 015013
  [arXiv:hep-ph/0507283].

\bibitem{g-2exp2}Bennett G et al., {\it Final report of the E821 muon anomalous magnetic moment 
measurement at BNL}, 2006 {\it Phys. Rev. D} {\bf 73} 072003 [hep-ex/0602035]
\bibitem{DDDD}Davier M, {\it The Hadronic Contribution to g}, 2007 {\it Nucl. Phys. B Proceedings S.} 
{\bf 169} 288 [hep-ph/0701163]
\bibitem{g-2MSSMf1l}Moroi T, {\it Muon anomalous magnetic dipole moment in the minimal supersymmetric 
standard model}, 1996 {\it Phys. Rev. D} {\bf 53} 6565 [hep-ph/9512396]
See also Moroi T, {\it Erratum: Muon anomalous magnetic dipole moment in the minimal supersymmetric 
standard model [Phys. Rev. D 53, 6565 (1996)]}, 1997 {\it Phys. Rev. D} {\bf 56} 4424
\bibitem{g-2CNH}Heinemeyer S, St\"ockinger D and Weiglein G, {\it Electroweak and supersymmetric two-loop 
corrections to $(g-2)_\mu$}, 2004 {\it Nucl. Phys. B} {\bf 699} 103 [hep-ph/0405255]
\bibitem{Batt}Battaglia M, Roeck A D, Ellis J, Gianotti F, Olive K A and Pape L, {\it Updated post-WMAP 
benchmarks for supersymmetry} 2004 {\it European Physical J. C} {\bf 33} 273 [hep-ph/0306219]
\bibitem{Higgs}ALEPH Collaboration, DELPHI Collaboration, L3 Collaboration, OPAL Collaboration
and The LEP  Working Group For Higgs Boson Searches, {\it Search for the Standard Model Higgs boson at LEP} 
2003 {\it Phys. Lett. B} {\bf 565} 61 [hep-ex/0306033]
\bibitem{:2001xx}ALEPH Collaboration, DELPHI Collaboration, L3 Collaboration, OPAL Collaboration
and the LEP Higgs Working Group, {\it Searches for the Neutral Higgs Bosons of the MSSM: Preliminary 
Combined Results Using LEP Data Collected at Energies up to 209 GeV} 2001 {\it Submitted to EPS'01 
in Budapest and Lepton-Photon '01 in Rome} [hep-ex/0107030]
\bibitem{Olechowski:1988gh}Olechowski M and Pokorski S, {\it Hierarchy of quark masses in the isotopic 
doublets in N=1 supergravity models} 1988 {\it Pjys. Lett. B} {\bf 214} 393
\bibitem{Ananthanarayan:1992cd}Ananthanarayan B, Lazarides G and Shafi Q, {\it Radiative electroweak 
breaking and sparticle spectroscopy with tan$\beta\simeq m_t/m_b$} 1993 {\it Phys. Lett. B} 
{\bf 300} 245
\bibitem{Djouadi:2004ae}Djouadi A, Heinemeyer S, Mondragon M and Zoupanos G, {\it Finite Unified Theories
and the Higgs Mass Prediction} 2004 {\it Springer Proceedings in Physics} {\bf 98} 273 [hep-ph/0404208]
\bibitem{Kapetanakis:1992vx}Kapetanakis D, Mondragon M and Zoupanos G, {\it Finite unified models} 1993 
{\it Zeitschrift f\"ur Physik C Particles and Fields} {\bf 60} 181 [hep-ph/9210218]
\bibitem{Yao:2006px}Yao W M et al. (Particle Data Group), {\it Review of Particle Physics} 2006 
{\it J. Phys. G} {\bf 33} 1 [astro-ph/0601514]
\bibitem{Feng} Feng J L, {\it Dark matter at the Fermi scale} 2006, {\it Journal of Physics G: Nuclear and 
Particle Physics} {\bf 32} R1 [astro-ph/0511043]
\bibitem{CDMS} CDMS collaboration, {\it A Search for WIMPs with the First Five-Tower Data from CDMS} 
2008 [arXiv:0802.3530]
\bibitem{eke_ent}Eke V R, Navarro J F and Frenk C S, {\it The Evolution of X-Ray Clusters in a 
Low-Density Universe} 1998 {\it Astrophys. J.} {\bf 503} 569 [astro-ph/9708070]
\bibitem{entropia_halos}Faltenbacher A, Hoffman Y, Gtlober S and Yepes G, {\it Entropy of gas and dark 
matter in galaxy clusters} 2007 {\it Mon. Not. R. Astron. Soc.} {\bf 376} 1327 [astro-ph/0608304]
\bibitem{NFW}Navarro J F, Frenk C S and White S D M, {\it A Universal Density Profile from Hierarchical 
Clustering} 1997 {\it Astrophys. J.} {\bf 490} 493 [astro-ph/9611107]
\bibitem{Nata07}Natarajan P, Croton, D and Bertone G, {\it Consequences of dark matter self-annihilation 
for galaxy formation} 2007 [arXiv:0711.2302]
\bibitem{Mo}Mo H J, Mao S and White S D M, {\it The formation of galactic discs} 1998 {\it Mon. Not. R.
Astron. Soc.} {\bf 295} 319 [astro-ph/9707093]
\bibitem{Lokas}Lokas E L and Mamon G A, {\it Properties of spherical galaxies and clusters with an NFW 
density profile} 2001 {\it Mon. Not. R. Astron. Soc.} {\bf 321} 155 [astro-ph/0002395]
\bibitem{Lokas2}Lokas E L and Hoffman Y, {\it Nonlinear evolution of spherical perturbation in a non-flat 
Universe with cosmological constant} 2001 [astro-ph/0108283]
\bibitem{Lokas3}Lokas E L, {\it Structure Formation in the Quintessential Universe} 2001 
{\it Acta Phys. Pol. B} {\bf 32} 3643
\bibitem{Neto}Neto A F, Gao L, Bett P, Cole S, Navarro J F, Frenk C S, White S D M, Springel V and Jenkins A,
{\it The statistics of $\Lambda$CDM halo concentrations} 2007 {Mon. Not. R. Astron. Soc.} {\bf 381} {1450}
\bibitem{MS}Springel V et al., {\it Simulations of the formation, evolution and clustering of galaxies and quasars}
2005 {\it Nature} {\bf 435} 629
\bibitem{Zavala}Zavala J, Avila-Reese V, Hern\'andez-Toledo H and Firmani C, {\it The luminous and dark 
matter content of disk galaxies} 2003 {\it Astron. Astrophys.} {\bf 412} 633 [astro-ph/0305516]
\bibitem{Blok1}de Blok W J G, McGaugh S S and Rubin V C, {\it High-Resolution Rotation Curves of Low 
Surface Brightness Galaxies. II. Mass Models} 2001 {\it Astron. J.} {\bf 122} 2396
\bibitem{Blok2}de Blok W J G, McGaugh S S, Bosma A and Rubin V C, {\it Mass Density Profiles of Low 
Surface Brightness Galaxies} 2001 {\it Astrophys. J.} {\bf 552} 23 [astro-ph/0103102]
\bibitem{Bin}Binney J J and Evans N W, {\it Cuspy dark matter haloes and the Galaxy} 2001 
{\it Mon. Not. R. Astron. Soc.} {\bf 327} L27 [astro-ph/0108505]
\bibitem{B-S}Borriello A and Salucci P, {\it The dark matter distribution in disc galaxies} 2001 
{\it Mon. Not. R. Astron. Soc.} {\bf 323} 285 [astro-ph/0001082]
\bibitem{B-O}Blais-Ouellette S, Carignan C and Amram P, {\it Multiwavelength Rotation Curves to Test Dark 
Halo Central Shapes} 2002 {\it ASP Conference Proceedings} {\bf 282} 129 [astro-ph/0203146]
\bibitem{Bosma3}Bosma A, {\it Dark Matter in Galaxies: Observational overview} 2004 
{\it IAU Symposium 220, Sydney, Australia. Eds: S. D. Ryder, D. J. Pisano, M. A. Walker, and K. C. Freeman.
San Francisco: Astronomical Society of the Pacific., p.39} [astro-ph/0312154]
\bibitem{Pointe}Pointecouteau E, Arnaud M and Pratt G W, {\it The structural and scaling properties of 
nearby galaxy clusters. I. The universal mass profile} 2005 {\it Astron. Astrophys.} {\bf 435} 1 
[astro-ph/0501635]
\bibitem{Voigt}Voigt L M and Fabian A C, {\it Galaxy cluster mass profiles} 2006 {\it Mon. Not. R. Astron. 
Soc.} {\bf 368} 518 [astro-ph/0602373]
\bibitem{Zapa}Zappacosta L, Buote D A, Gastaldello F, Humphrey P J, Bullock J, Brighenti F and
Mathews W, {\it The Absence of Adiabatic Contraction of the Radial Dark Matter Profile in the Galaxy 
Cluster A2589} 2006 {\it Astrophys. J.} {\bf 650} 777 [astro-ph/0602613]
\bibitem{Nav04}Navarro J F, Hayashi E, Power C, Jenkins A R, Frenk C S, White S D M, Springel V, Stadel J and
Quinn T R, {\it The inner structure of $\Lambda$CDM haloes - III. Universality and asymptotic slopes} 2004, 
{\it Mon. Not. R. Astron. Soc.} {\bf 349} 1039 [astro-ph/0311231]
\bibitem{Die05}Diemand J, Zemp M, Moore B, Stadel J, Carollo C M, {\it Cusps in cold dark matter haloes}
2005, {\it Mon. Not. R. Astron. Soc.} {\bf 364} 665 [astro-ph/0504215] 
\bibitem{Matos}Matos T, Nunez D and Sussman R A, {\it A general relativistic approach to the Navarro 
Frenk White galactic halos} 2004 {\it Class. Quantum Grav.} {\bf 21} 5275 [astro-ph/0410215]
\bibitem{Dehnen}Dehnen W and McLaughlin D, {\it Dynamical insight into dark matter haloes} 2005 
{\it Mon. Not. R. Astron. Soc.} {\bf 363} 1057 [astro-ph/0506528]
\bibitem{Moore}Hansen S H and Moore B, {\it A universal density slope Velocity anisotropy relation for 
relaxed structures} 2006 {\it New Astron.} {\bf 11} 333
\bibitem{Kuzio06}Kuzio de Naray R, McGaugh S S, de Blok W J G and Bosma A, {\it High-Resolution Optical Velocity Fields of 11 
Low Surface Brightness Galaxies} 2006 {\it Astrophys. J. S. S.} {\bf 165} 461 [astro-ph/0604576]
\bibitem{Firmani}Firmani C, D'Onghia E, Chincarini G, Hern\'andez X and  Avila-Reese V, 
{\it Constraints on dark matter physics from dwarf galaxies through galaxy cluster haloes} 2001 
{\it Mon. Not. R. Astron. Soc.} {\bf 321} 713 [astro-ph/0005001]
\bibitem{Wilki}Wilkinson M I, Kleyna J T, Evans N W, Gilmore G F, Irwin M J and Grebel E K, 
{\it Kinematically Cold Populations at Large Radii in the Draco and Ursa Minor Dwarf Spheroidal Galaxies}
2004 {\it Astrophys. J.} {\bf 611} L21
\bibitem{Prada}S\'anchez-Conde M A, Prada F, Lokas E L, G\'omez M E, Wojtak R and Moles M,
{\it Dark matter annihilation in Draco: New considerations of the expected gamma flux}
2007 {\it Phys. Rev. D} {\bf 76} 123509 [astro-ph/0701426] 
\bibitem{Lokas_sig1}Lokas E L, Mamon G A and Prada F, {\it Dark matter distribution in the Draco dwarf from velocity 
moments} 2005 {\it Mon. Not. R. Astron. Soc.} {\bf 363} 918
\bibitem{Lokas_sig2}Mamon G A and Lokas E L, {\it Dark matter in elliptical galaxies - II. Estimating the mass 
within the virial radius} 2005 {\it Mon. Not. R. Astron. Soc.} {\bf 363} 705 
\bibitem{Ode}Odenkirchen M et al., {\it New Insights on the Draco Dwarf Spheroidal Galaxy from the Sloan Digital 
Sky Survey: A Larger Radius and No Tidal Tails} 2001 {\it Astrophys. J.} {\bf 122} 2538 
\bibitem{Walker07} Walker M G, Mateo M, Olszewski E W, Gnedin O Y, Wang X, Sen B and Woodroofe M, 
{\it Velocity Dispersion Profiles of Seven Dwarf Spheroidal Galaxies} 2007 {\it  Astrophys. J. L.} {\bf 667} L53 
\end{thebibliography}
\end{document}